\documentclass[reprint,preprintnumbers,amsmath,amssymb,aps,prd,superscriptaddress,nofootinbib,floatfix]{revtex4-2}

\usepackage{graphicx}
\usepackage{dcolumn}
\usepackage{bm}
\usepackage{hyperref}
\usepackage[mathlines]{lineno}
\usepackage{siunitx}
\usepackage{xspace}

\newcommand{\ddjj}{$\delta\Delta_{\mathrm{JJ}}$\xspace}
\newcommand{\ddgnd}{$\delta\Delta_{\mathrm{gnd}}$\xspace}

\usepackage{xcolor}
\usepackage{multirow}
\usepackage{booktabs}

\begin{document}

\preprint{}


\title{Distinguishing types of correlated errors in superconducting qubits}

\newcommand{\rle}{\affiliation{Research Laboratory of Electronics, Massachusetts Institute of Technology, Cambridge, MA 02139, USA}}
\newcommand{\lns}{\affiliation{Laboratory for Nuclear Science, Massachusetts Institute of Technology, Cambridge, MA 02139, USA}}
\newcommand{\lincoln}{\affiliation{Lincoln Laboratory, Massachusetts Institute of Technology, Lexington, MA 02421, USA}}
\newcommand{\jhuapl}{\affiliation{Johns Hopkins University Applied Physics Laboratory, Laurel, MD 20723, USA}}
\newcommand{\phys}{\affiliation{Department of Physics, Massachusetts Institute of Technology, Cambridge, MA 02139, USA}}
\newcommand{\eecs}{\affiliation{Department of Electrical Engineering and Computer Science, Massachusetts Institute of Technology, Cambridge, MA 02139, USA}}

\author{H.~P.~Binney} \email{hbinney@mit.edu} \rle\lns
\author{H.~D.~Pinckney}  \rle\lns
\author{K.~Azar} \rle\eecs
\author{P.~M.~Harrington} \rle\lns
\author{S.~Jha} \rle\eecs
\author{M.~Li} \lns\phys
\author{J.~Yang} \lns\phys
\author{F.~Contipelli} \lincoln
\author{R.~DePencier Piñero} \lincoln
\author{M.~Gingras} \lincoln
\author{B.~M.~Niedzielski} \lincoln
\author{H.~Stickler} \lincoln
\author{M.~E.~Schwartz} \lincoln
\author{J.~A.~Grover} \rle
\author{M.~Hays} \rle
\author{K.~Serniak} \rle\lincoln
\author{J.~A.~Formaggio} \lns\phys
\author{W.~D.~Oliver} \rle\eecs\phys

\date{\today}

\begin{abstract}
Errors in superconducting qubits that are correlated in time and space can pose problems for quantum error correction codes.  Radiation from cosmic and terrestrial sources can increase the quasiparticle (QP) density in a superconducting qubit device, resulting in an increased rate of QPs tunneling across proximal Josephson junctions (JJs) and causing correlated errors.  Mechanical vibrations, such as those induced by the pulse tube (PT) in a dry dilution refrigerator, are also a known source of correlated errors.  We measure two types of errors in the same device, linking the first to ionizing radiation and the second to PT operation.  We present a method for distinguishing these two types of errors by their temporal, spatial, and frequency domain features, enabling physically motivated error-mitigation strategies.  We also present accelerometer data to study the correlation between PT-induced vibrations and the errors.  We measure arrays of transmon qubits where the difference in superconducting gap across the JJ is less than the qubit energy, as well as those where the gap is greater than the qubit energy, which has been shown to mitigate radiation-induced errors.  The rate of both types of errors is reduced in these latter devices, suggesting that gap engineering is also protective against PT-induced errors.
\end{abstract}
\maketitle


\section{\label{sec:introduction}Introduction}

\begin{figure}
    \centering
    \includegraphics[width=\linewidth]{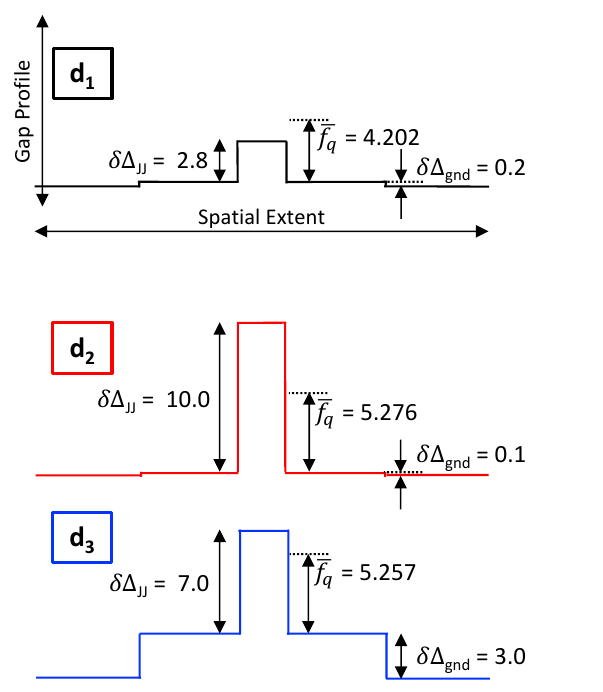}
    \caption{The superconducting gap profile of the three devices ($\mathrm{d_1}$, $\mathrm{d_2}$, and $\mathrm{d_3}$) under test.  The vertical axis of each profile represents the superconducting gap, while the horizontal axis represents the spatial extent along the JJ.  The \ddjj parameter refers to the difference in superconducting gap across the JJ.  When \ddjj is greater than the qubit frequency $f_{\mathrm{q}}$, quasiparticle tunneling across the JJ is inhibited. The \ddgnd parameter refers to the difference in superconducting gap between the JJ and the ground plane.  Values of \ddjj, \ddgnd, and the average qubit frequency $\bar{f_{\mathrm{q}}}$ are labeled for each device, in units of GHz.  The values of $\bar{f_{\mathrm{q}}}$ were measured, whereas \ddjj and \ddgnd are the designed values, estimated from transition temperature measurements of aluminum films from the same fabrication process.  Reproduced in part from Ref.~\protect{\onlinecite{pinckney2026characterizationradiationinducederrorssuperconducting}}.}
    \label{fig:gap_profile}
\end{figure}
Superconducting qubits are a promising candidate for realizing quantum computers.  Over the past six years, radiation from cosmic rays and terrestrial sources has been identified as a source of qubit decoherence \cite{vepsalainenImpactIonizingRadiation2020, wilenCorrelatedChargeNoise2021, thorbeckTwoLevelSystemDynamicsSuperconducting2023, harringtonSynchronousDetectionCosmic2024, liDirectEvidenceCosmicrayinduced2024a, bratrudFirstMeasurementCorrelated2024a, dominicisEvaluatingRadiationImpact2024,nhoRecoveryDynamicsGapengineered2025,kurilovichCorrelatedErrorBursts2025}.  High energy particles impacting the device and substrate create quasiparticles (QPs), which cause qubit errors when they tunnel across the Josephson junctions (JJs).  This error mechanism can persist for extended periods of time (10s of milliseconds) and can simultaneously affect many proximal qubits \cite{acharyaSuppressingQuantumErrors2023, yeltonModelingPhononmediatedQuasiparticle2024, harringtonSynchronousDetectionCosmic2024}.  These spatially and temporally correlated errors pose problems for traditional quantum error correction codes, which assume that errors are sufficiently uncorrelated \cite{acharyaQuantumErrorCorrection2024, tanResilienceSurfaceCode2024}.

Environmental effects aside from radiation can also induce correlated qubit errors.  Qubit devices are commonly operated in dry dilution refrigerators (DRs) which are usually precooled by a pulse tube cryocooler (PT).  The PT induces periodic vibrations, which have been associated with correlated errors in superconducting qubits \cite{kono_mechanically_2024} and electronic noise caused by triboelectric effects \cite{kalra_vibration-induced_triboelectric_2016}.  Vibrations are also a common challenge for cryogenic detectors used in fundamental physics experiments, with some implementing extensive vibration isolation infrastructure and active noise reduction \cite{cuore_vibrations, cuore_active_noise_cancel, pirro_cryo_vibratioN_Reduction}, as well as noise-cancellation algorithms using accelerometers and other auxiliary sensors \cite{vetter_improving_2024}.  


Understanding the source of correlated qubit errors informs mitigation techniques.  QP-induced errors can be mitigated either by reducing sources of QPs or by reducing the qubits' sensitivity to them.  For example, quantum computers can be shielded from cosmic rays by operating in an underground facility \cite{cardani_reducing_2021, bratrudFirstMeasurementCorrelated2024a}, which reduces QPs induced by ionizing radiation.  However, substantial overburden is required to reduce cosmic ray muon flux, and these facilities do not mitigate radiation sources inherent in the local environment; this requires additional shielding \cite{loerAbatementIonizingRadiation2024, dominicisEvaluatingRadiationImpact2024}.  Vibrations could be mitigated by operating in a ``wet" DR with no PT \cite{kalra_vibration-induced_triboelectric_2016}, using vibration isolation and cancellation strategies, or by operating for short periods of time with the PT turned off \cite{yelton-h65v-ttbw}.  Large-scale cryogenic facilities that do not use PTs would also mitigate PT-induced errors, but this would require infrastructure not accessible in many laboratory settings \cite{croot2025enablingtechnologiesscalablesuperconducting, Claudet2000CryoPlant}.

Due to challenges in reducing sources of QPs, a variety of on-chip mitigation strategies has been investigated to reduce the qubits' sensitivity to them.  Increasing the difference in the superconducting gap across the JJ so that it is greater than the qubit energy, commonly referred to as gap engineering, has been proven to reduce radiation-induced correlated errors \cite{aumentadoNonequilibriumQuasiparticles2e2004, mcewenResistingHighEnergyImpact2024, nhoRecoveryDynamicsGapengineered2025,kurilovichCorrelatedErrorBursts2025, harringtonPatent, pinckney2026characterizationradiationinducederrorssuperconducting, diamondDistinguishingParitySwitchingMechanisms2022} but has not been tested in the case of PT-induced errors.

In this work, we measured two populations of correlated errors in the same device.  We attributed one type to ionizing radiation, given past measurements in this and similar devices \cite{harringtonSynchronousDetectionCosmic2024, pinckney2026characterizationradiationinducederrorssuperconducting}.  We linked the other to PT operation, hypothesizing that the vibrations induced by the PT are the source of these correlated errors.  We determined the spatial, temporal, and frequency domain features of each type, enabling us to identify and distinguish them.  The second population of errors vanished when the PT was turned off.  We also studied the correlation between vibrations and this second population of errors by operating in two different DRs, each of which had a different mass and structural configuration and therefore a different vibration environment, as confirmed with accelerometers.  We did not observe PT-induced correlated errors in the second DR, which likely contributed to improved coherence times.  Finally, we measured a reduction in both radiation- and PT-induced errors in gap-engineered devices.

\section{\label{sec:devices} Qubit devices and data acquisition}


Over the course of the experimental campaign, we acquired data from 3 devices, each with 10 superconducting fixed-frequency transmon qubits (the same devices measured in Ref.~\onlinecite{pinckney2026characterizationradiationinducederrorssuperconducting}).  The devices were exposed to cosmic and terrestrial radiation naturally existing in the laboratory environment, as characterized in Refs.~\onlinecite{vepsalainenImpactIonizingRadiation2020} and \onlinecite{harringtonSynchronousDetectionCosmic2024}.  The superconducting gap profile of the three devices is shown in Fig.~\ref{fig:gap_profile}.  The first device, referred to as device $\mathrm{d_1}$, had an average qubit frequency $\bar{f_{\mathrm{q}}}$ greater than the gap difference across the Josephson junction, \ddjj.  This device had the same design tested in Ref.~\onlinecite{harringtonSynchronousDetectionCosmic2024}, although it was not the same device.  The other two devices were fabricated to have \ddjj greater than $\bar{f_{\mathrm{q}}}$, providing protection against QPs tunneling across the JJ.  The second device, $\mathrm{d_2}$, had the largest \ddjj ($\approx$ 10.0 GHz).  The third device, $\mathrm{d_3}$, had a lower \ddjj ($\approx$ 7.0 GHz) but a larger gap difference between the JJ leads and the ground plane \ddgnd ($\approx$ 3.0 GHz).  The larger \ddgnd has also been shown to mitigate radiation-induced errors \cite{pinckney2026characterizationradiationinducederrorssuperconducting}\footnote{Device $\mathrm{d_1}$ corresponds to the ``no blocking" device in Ref.~\onlinecite{pinckney2026characterizationradiationinducederrorssuperconducting}, $\mathrm{d_2}$ corresponds to ``JJ only" device, and $\mathrm{d_3}$ corresponds to the ``JJ and M1" device (M1 indicates an aluminum base metal layer). The average qubit frequency across the array slightly differs from Ref. \cite{pinckney2026characterizationradiationinducederrorssuperconducting} due to JJ aging effects, as the measurements for each paper were taken in distinct DR cool downs.}

\begin{figure}[ht]
    \centering
    \includegraphics[width=\linewidth]{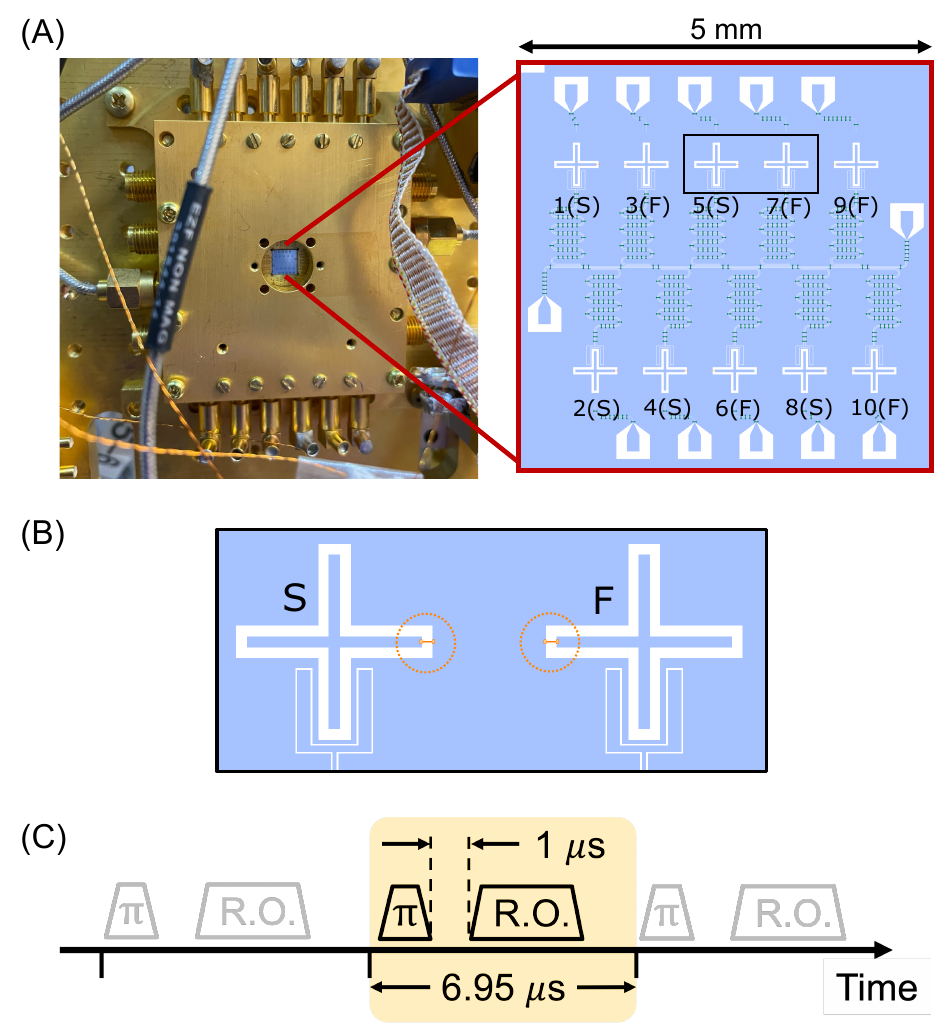}
    \caption{Details on the device mounting and measurement sequence.  (A) A picture of device $\mathrm{d_1}$ mounted in a DR.  The device package is mounted vertically, and the device is suspended by wirebonds from the package.  The panel on the right shows an artificially colored layout of the qubit array, reproduced from Ref.~\protect{\cite{pinckney2026characterizationradiationinducederrorssuperconducting}}, with the 10 qubits arranged in two rows of 5.  The qubit number and junction orientation (S for slow recovery or F for fast recovery) are labeled.  (B) A closeup of the black box in panel (A), showing the location of the JJ (in orange) for the two orientations.  (C) The readout sequence for all measurements.  A $\pi$-pulse is applied to change the qubit state.  After 1 \unit{\micro\second}, the qubit is read out.  The cadence of the full measurement is 6.95 \unit{\micro\second}.}
    \label{fig:devices}
\end{figure}

Fig.~\ref{fig:devices}(A) shows a qubit device mounted in a DR.  Devices were mounted vertically (edges of the chip oriented up, down, left, and right) on a cold finger that was affixed perpendicularly to the mixing chamber (MXC) plate.  All three devices were suspended by aluminum wedge wire bonds and were not glued into place, with the goal of avoiding low-energy phonon-mediated bursts which have been associated with mechanical stress in glue \cite{anthony-petersen_stress-induced_2024, yelton-h65v-ttbw} and to avoid placing a lossy material near the qubits.  However, this choice may have increased the devices' sensitivity to vibrations.

In each device, the qubits were arranged in two offset rows, as shown in Fig.~\ref{fig:devices}(A, right).  In addition, each device had 5 qubits with one of two junction orientations, corresponding to whether the JJ lead with the higher superconducting gap was adjacent to the qubit capacitor or the ground plane, as shown in Fig.~\ref{fig:devices}(B).  
As shown in Refs.~\onlinecite{harringtonSynchronousDetectionCosmic2024} and \onlinecite{pinckney2026characterizationradiationinducederrorssuperconducting}, the junction orientation affects the QP dynamics at the JJ, resulting in slower or faster recovery times, respectively, after an ionizing radiation event.  In these devices, qubits 1, 2, 4, 5, and 8 were orientation ``S," which had a slower recovery time from radiation-induced errors ($\approx 5$ \unit{ms}), and qubits 3, 6, 7, 9, and 10 were orientation ``F," which had a faster recovery time ($\approx 0.7$ \unit{ms}) \cite{harringtonSynchronousDetectionCosmic2024}.

We tested device $\mathrm{d_1}$ in two different dry DRs referred to as DR1 and DR2.  DR1 is a Leiden CF-CS81-1500 dilution refrigerator with a Cryomech PT415-RM pulse tube, and DR2 is a Bluefors XLD-400 dilution refrigerator with a Cryomech PT425-RM pulse tube.  The two DRs were in different buildings, possibly resulting in different radiation environments. The difference in mass and PT mounting between the two experimental setups, and the implications on the correlated qubit errors, is discussed further in Section~\ref{sec:accel}.  We measured the remaining two devices only in DR1.  

We repeatedly  measured the qubit using the sequence shown in Fig.~\ref{fig:devices}(C), which is primarily sensitive to relaxation errors \cite{harringtonSynchronousDetectionCosmic2024, pinckney2026characterizationradiationinducederrorssuperconducting}.  A $\pi$-pulse was applied to change the state of the qubit, and the qubit state was read out after a 1 \unit{\micro s} delay.  A qubit relaxation occurred when a qubit was prepared into the excited state and measured in the ground state.  If the qubit was in the excited state prior to the $\pi$-pulse, as determined by the previous measurement, then the measurement was discarded.  The presence of any non-quantum non-demolition (non-QND) measurement effects would decrease sensitivity to qubit relaxation and equally affect each source of correlated errors.  We define $N_{\text{r}, i}$ to be the number of relaxations (0 or 1) in each qubit $i$ at each measurement.  As QPs tunnel across the JJ, they induce additional qubit relaxations beyond the baseline rate, which can be measured with this sequence.  Each data file comprised one million measurements of $N_{\text{r}, i}$, for a total measurement time of 6.95 seconds per file.  Each of the datasets shown in Fig.~\ref{fig:filter_vs_decay} comprised between 999 and 2500 files, corresponding to between 1.9 and 4.9 hours of total data acquisition duration.  The duration of each measurement is shown in Table \ref{tab:dataset_times}.

\begin{table}[]
    \centering
    \centering
    \begin{tabular}{c c c}
        \toprule
        DR           & \hspace{0.5cm} Device  &  \hspace{0.5cm} Duration [hrs] \\
        \midrule
        DR1  & \hspace{0.5cm} $\mathrm{d_1}$ &  \hspace{0.5cm} 4.8 \\
        DR2 & \hspace{0.5cm} $\mathrm{d_1}$ &  \hspace{0.5cm} 2.5 \\
        DR1 & \hspace{0.5cm} $\mathrm{d_2}$&  \hspace{0.5cm} 4.8 \\
        DR1 & \hspace{0.5cm} $\mathrm{d_3}$ & \hspace{0.5cm} 1.9\\

        \bottomrule
    \end{tabular}
    \caption{Duration of the four datasets compared in Fig.~\ref{fig:filter_vs_decay}.}
    \label{tab:dataset_times}
\end{table}

\section{\label{sec:distinguishing} Distinguishing types of correlated qubit errors}
\begin{figure}[h!]
    \centering
    \includegraphics[width=\linewidth]{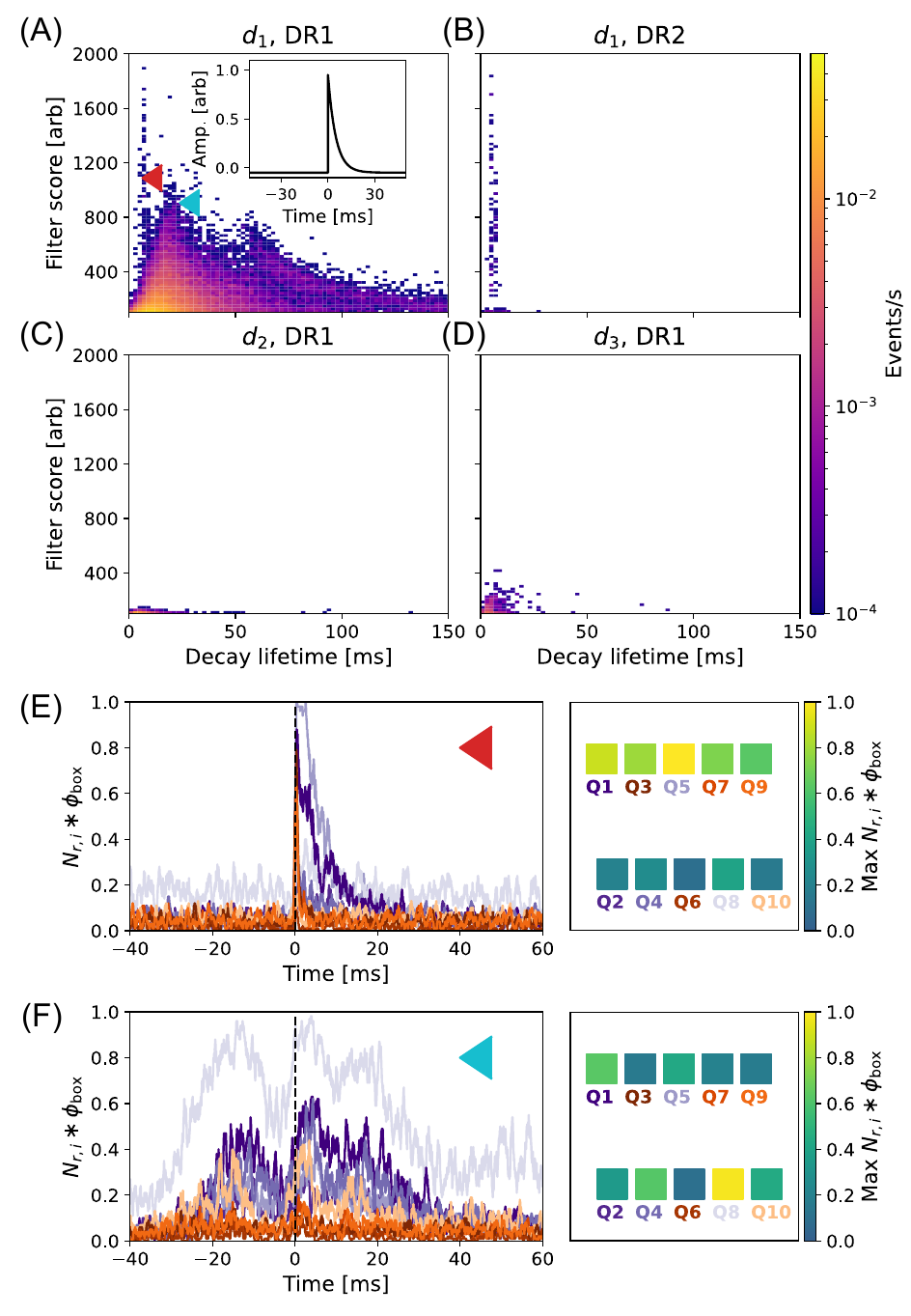}
    \caption{Distinguishing the two types of correlated qubit errors.  (A) A histogram of filter score vs. decay lifetime for all error events detected in device $\mathrm{d_1}$ in DR1.  The exponential filter used to identify error event candidates within the $N_\text{r}$ data is inset.  The filter has an exponential decay lifetime of $\tau_0$ = 5 \unit{ms}, the expected recovery lifetime for S orientation qubits after a radiation-induced event \protect{\cite{harringtonSynchronousDetectionCosmic2024}}.  This filter is time-reversed before convolving with $N_\text{r}$.  Multiple error populations are visible.  The red marker points to the filter score and decay lifetime of the radiation-induced event in (E), and the cyan marker points to the PT-induced event in (F).  (B) Histogram of filter score vs. decay lifetime for device $\mathrm{d_1}$ in DR2, showing a reduction in PT-induced events.  (C) Histogram of filter score vs. decay lifetime for device $\mathrm{d_2}$, showing a reduction in all events.  (D) Histogram of filter score vs. decay lifetime for device $\mathrm{d_3}$, showing a reduction in all events. (E) A sample radiation-induced event.  The left panel shows a moving average of the number of relaxations measured in each qubit, $N_{\text{r}, i} \ast \phi_{\mathrm{box}}$, for a radiation-induced error.  The dashed line corresponds to the start time of the event.  The right panel shows the maximum value of $N_{\text{r}, i} \ast \phi_{\mathrm{box}}$ for each of the qubits.  S orientation qubits are shown in purple colors, and F orientation qubits are shown in orange colors.  (F) A sample PT-induced event.}
    \label{fig:filter_vs_decay}
\end{figure}

We measured device $\mathrm{d_1}$ in DR1 for about 5 hours.  In order to identify correlated error event candidates, we first summed $N_{\text{r}, i}$ across the 10 qubits to obtain $N_\text{r} = \sum_{i} N_{\text{r}, i}$, where $i$ corresponds to the qubit number.  Then, $N_\text{r}$ was convolved with the time-reversed version of an exponential filter, referred to as $\phi_{\mathrm{exp}}$ (this procedure is commonly known as matched filtering).  The filter before time-reversal is shown in Fig.~\ref{fig:filter_vs_decay}(A, inset).  The filter had an exponential decay lifetime of $\tau_0 = 5$ \unit{ms}, corresponding to the approximate decay lifetime of a radiation-induced error burst as measured previously in this device \cite{harringtonSynchronousDetectionCosmic2024}.   We describe the matched filter procedure in more detail in Appendix~\ref{sec:matched_filtering}.

Candidate error events were identified based on a peak-finding algorithm in the convolved sequence $C = N_\text{r} \ast \phi_{\mathrm{exp}}$.  The ``filter score" is defined as the maximum value of $C$ for each candidate event.  This filter score gives information about the magnitude of the error event as well as its resemblance to the shape of the filter.  The start time of the event is defined as the time at which the filter score reached a maximum.  We define the filter function so that it integrates to zero, meaning that the filter score is distributed around zero in the absence of an error event, regardless of the baseline error rate.  This makes it sensitive to the change in the relaxation rate, rather than the absolute rate.

After identifying error event candidates, $N_\text{r}$ was also convolved with a boxcar filter (moving average) $\phi_{\mathrm{box}}$ of length 100 measurements for visualization and fitting.  An exponential decay fit was applied to $N_\text{r} \ast \phi_{\mathrm{box}}$ after the start time to extract the decay lifetime of the error event.  The decay lifetime $\tau$ was allowed to float, while the $\tau_0 = 5$ \unit{ms} lifetime for the filter was kept constant when identifying error candidates and calculating the filter score.

Fig.~\ref{fig:filter_vs_decay}(A) shows the distribution of the filter score versus decay lifetime for event candidates in device $\mathrm{d_1}$ measured in DR1.  We note that subplots (B), (C), and (D) are provided here for visual comparison; they are discussed in detail in Section~\ref{sec:fridge2} and \ref{sec:gapEng}.  Significant structure in the event distribution is apparent.  Events that have a short decay lifetime and large filter score, as seen in the upper left of the figure, are the expected population of events arising from cosmic rays and terrestrial radiation \cite{harringtonSynchronousDetectionCosmic2024, pinckney2026characterizationradiationinducederrorssuperconducting}.  These correlated qubit errors have a decay lifetime of about 5 ms, have a high filter score, and resemble the exponential decay template used to identify them.  

An example of a radiation-induced event is shown in Fig.~\ref{fig:filter_vs_decay}(E); the red marker in Fig.~\ref{fig:filter_vs_decay}(A) points to the filter score and decay lifetime parameters corresponding to this event.  In the left panel, the y axis is the running 100-measurement average of the number of relaxations in each qubit, $N_{\text{r}, i} \ast \phi_{\mathrm{box}}$.  Therefore, 1 is the maximum value (100\% probability of relaxation), and Q5 is saturating, meaning that the qubit is having the maximum possible response to the error event for an extended period.  The vertical dashed line indicates the start time.  The distinctive exponential decay shape is visible.  Additionally, the inhomogeneity of the recovery times between S (in purple colors) and F (in orange colors) orientations can be seen.  Qubits Q3, Q7, and Q9, which have the F orientation, have a much shorter recovery time than Q1 and Q5, which have the S orientation.  The panel on the right shows the maximum value of $N_{\text{r}, i} \ast \phi_{\mathrm{box}}$ around the event start time.  We note that the magnitudes are not directly comparable between qubits of type S and F, and that Q8 generally has a higher baseline and larger response than the others across events.  While each radiation-induced error burst is a combination of slow and fast responses, the sum across qubits is generally dominated by the slow response lifetime, justifying the choice of $\tau_0 = 5$ \unit{ms}.  Using a filter with a 0.7 ms lifetime, for example, still finds error event candidates with a lifetime of $\tau \approx 5$ \unit{ms}.

In addition to radiation-induced errors, another population of events is evident from the distribution in Fig.~\ref{fig:filter_vs_decay}(A).  A characteristic example of such an event is shown in Fig.~\ref{fig:filter_vs_decay}(F), corresponding to the cyan marker in Fig.~\ref{fig:filter_vs_decay}(A).  These events have a much longer decay lifetime and can also be quite large in magnitude.  They have a longer rise time and a characteristic multi-peaked structure.  These events constitute the vast majority of error events ($\approx 99\%$) detected with this experimental setup.  

\begin{figure}
    \centering
    \includegraphics[width=\linewidth]{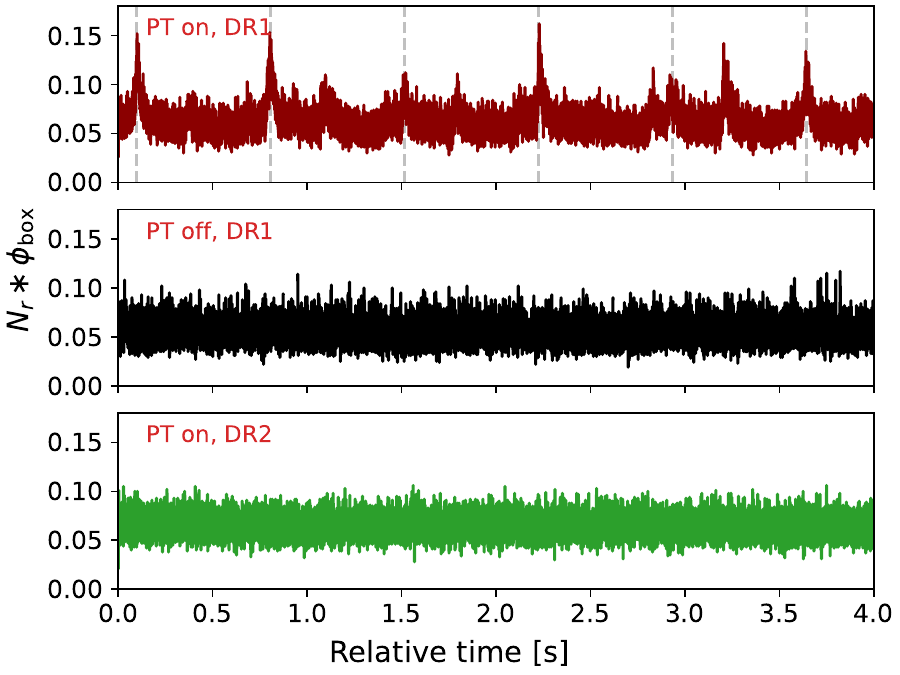}
    \caption{Number of relaxations averaged over 100 measurements as a function of time for device $\mathrm{d_1}$ in three measurement configurations.  In the top panel (PT on, DR1) the PT frequency is clearly visible, corresponding to the gray dashed lines.  In the other two panels, the PT frequency is not visible.  The x axis is relative time; these measurements were  not taken simultaneously.}
    \label{fig:relax_ts}
\end{figure}

Other studies have shown that the vibrations from the PT can cause correlated qubit errors and disrupt qubit coherence \cite{kono_mechanically_2024}.  If this second population of events is related to the PT, they should appear at the $\approx$ 1.4 Hz PT fundamental frequency.  Fig.~\ref{fig:relax_ts} shows $N_{\text{r}} \ast \phi_{\mathrm{box}}/10$ (relaxations averaged across the 10 qubits) as a function of time.  The top panel corresponds to a subset of the data in Fig.~\ref{fig:filter_vs_decay}(A), and the dashed lines correspond to the PT frequency.  These events clearly occur at the PT frequency, although the magnitude of the error is not the same each cycle.

We additionally quantify the degree of localization of the errors in the two populations.  We expect the radiation-induced errors to have a higher response in one area of the device where the radiation impacted the substrate, whereas a PT-induced error may not be similarly localized.  The qubits have highly inhomogeneous responses, and we have no independent knowledge of the location or energy of the radiation impacts.  These factors make it difficult to establish an accurate calibration that would allow us to determine the location of each event on the device.  Therefore, we define a simple metric $A$ which is insensitive to these inhomogeneities.  We define $A$ to be the difference between the sum of $\max(N_{\text{r}, i} \ast \phi_{\mathrm{box}})$ for the top ($i=1,3,5,7,9$) and the bottom ($i=2,4,6,8,10$) qubits of the array, over the average: 
\begin{equation}
    A = \frac{1}{2}\frac{\sum_{\mathrm{top}}\max(N_{\text{r}, i} \ast \phi_{\mathrm{box}}) - \sum_{\mathrm{bot.}}\max(N_{\text{r}, i} \ast \phi_{\mathrm{box}})}{\sum_{\mathrm{top}}\max(N_{\text{r}, i} \ast \phi_{\mathrm{box}}) + \sum_{\mathrm{bot.}}\max(N_{\text{r}, i} \ast \phi_{\mathrm{box}})}.
    \label{eq:A}
\end{equation}
\noindent When comparing the events in Fig.~\ref{fig:filter_vs_decay}(E) and (F), the event in panel (E) would have a greater value of $A$ due to the asymmetry in the response between the qubits at the top and the bottom of the device.  

We note that we could define $A$ for other combinations of qubits and get a similar result, since $A$ is simply constructed to show that the spatial response of the radiation-induced errors is significantly more varied than the PT-induced errors.  An example comparison for a different definition of $A$ is shown in Appendix~\ref{sec:app_A}.

\begin{figure}
    \centering
    \includegraphics[width=0.5\linewidth]{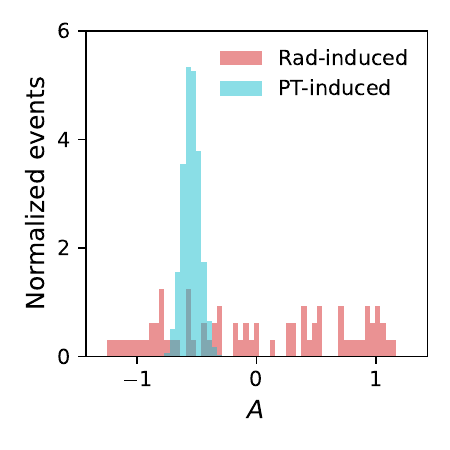}
    \caption{Histograms of the metric $A$ for the radiation- and PT-induced errors, showing that the location of the radiation-induced errors on the device varies across events, whereas the location of the PT-induced errors does not.  The negative bias in $A$ for the PT-induced events is likely due to the inhomogeneous response of the qubits across the device.  The area of each histogram integrates to one so that the shapes can be compared.}
    \label{fig:localization}
\end{figure}

Given that the event types have overlapping filter score and lifetime parameters, in order to compare values of $A$ between radiation- and PT-induced events, heuristic cuts are chosen to achieve relatively clean samples of each type.  We select a population of radiation-induced events by requiring a decay lifetime of less than 9 ms and a filter score of greater than 700.  We select PT-induced events by requiring a decay lifetime of greater than 12 ms and a filter score of greater than 300.  Fig.~\ref{fig:localization} shows histograms of $A$ for the two event populations.  The PT-induced events generally have a similar value of $A$, whereas the radiation-induced events vary in $A$.  This suggests that the spatial distribution for the PT-induced events is the same on average, whereas the radiation-induced events can be centered on any part of the device.  $A$ is likely negative for the PT-induced events because of the inhomogeneous response across qubits, such as the larger response in Q8, as seen in Fig.~\ref{fig:filter_vs_decay}(F).

In order to further study the connection between the PT and this second type of correlated qubit errors, we took several sets of measurements described in the following sections.

\section{\label{sec:PToff} Measurements with the pulse tube off}
To verify that the non-radiation-induced events were associated with the PT, we took data with the PT turned off.  The PT was originally turned off for two short periods of 30 \unit{s}.  Later, the PT and circulation were turned off for a longer period of about 8 minutes.  After 8 minutes, the temperatures at different DR stages began to rise, although we continued to take qubit data throughout the PT-off period and subsequent temperature recovery.  More details on the response of the DR to the PT-off measurements are provided in Appendix~\ref{sec:app_PToff}.  We note that the PT-on data in this section are different than those in the Section~\ref{sec:distinguishing}, where data were taken in normal conditions (PT on) for a longer period of time; however, similar behavior was seen in both datasets.

\begin{figure}
    \centering
    \includegraphics[width=\linewidth]{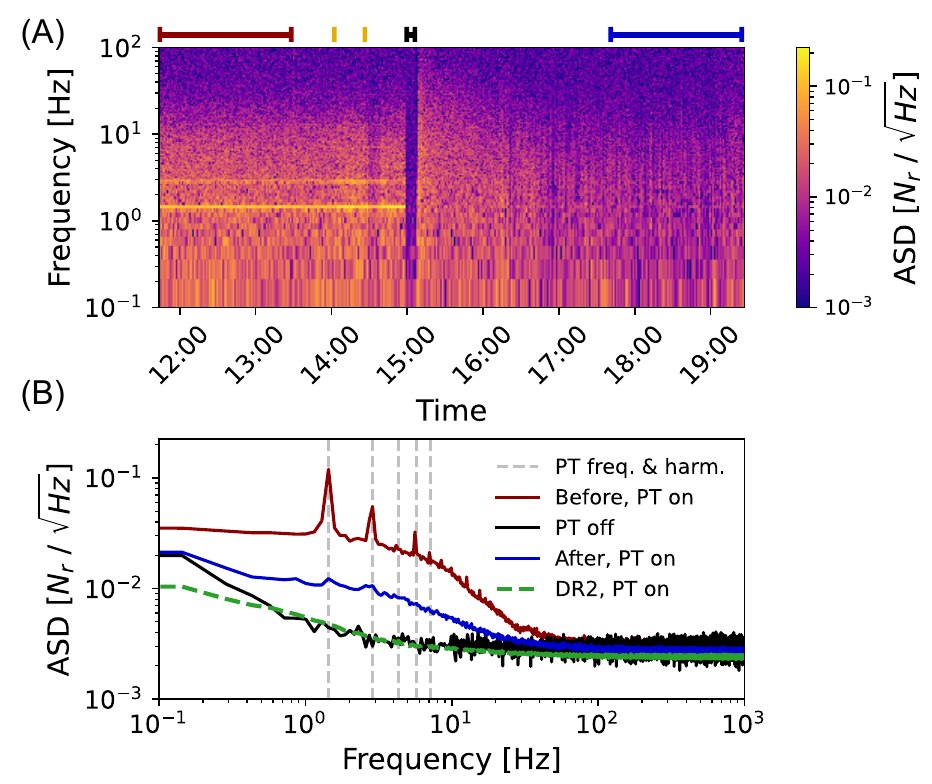}
    \caption{Amplitude spectral densities (ASDs) of $N_\text{r}$ with the PT on and off.  (A) Spectrogram measured with device $\mathrm{d_1}$ in DR1.  Before the PT was turned off (region marked above the panel, in red), a strong signal at the PT frequency and harmonics was visible.  The PT was turned off for brief periods around 14:00, marked in orange, and then a longer period around 15:00, marked in black.  The circulation was also turned off during this time period.  After the PT was turned back on, the ASD increased again but still had a lower magnitude than before.  The red, black, and blue portions of the spectrogram correspond to the averaged curves shown in (B).  (B) The averaged ASD curves for each of the regions highlighted in (A): with the PT on, with the PT off, and with the PT turned back on afterwards.  The green dashed line shows the ASD with the PT on in DR2, for comparison.}
    \label{fig:psds}
\end{figure}

Fig.~\ref{fig:psds}(A) shows a spectrogram of the amplitude spectral density (ASD) of the number of relaxations over a several hour measurement time.  Fig.~\ref{fig:psds}(B) shows the ASD from individual data acquisition periods, marked at the top of Fig.~\ref{fig:psds}(A), averaged over multiple 6.95 second long files.  At the beginning of the data acquisition period, corresponding to the red curve, the ASD has a large low-frequency feature, with prominent peaks at $\approx$ 1.4 Hz and harmonics.  Both radiation-induced and PT-induced correlated qubit errors will cause low-frequency features in the spectrum; however, this dataset contained many more PT-induced events than radiation-induced events.

The two short PT-off periods occurred around 14:00 (marked in orange), although they cannot easily be seen in the spectrogram.  The longer PT-off period is highlighted in black, around 15:00.  When the PT was turned off, the low-frequency feature immediately vanished, as shown in Fig.~\ref{fig:psds}(B) in black (this curve is less smooth than the others due to averaging fewer files).  This behavior suggests that the PT-induced events were the primary cause of the low-frequency features in the spectrum.  Additionally, the number of error candidate events found using the exponential filter immediately dropped; the corresponding time series with a portion of the PT-off data can be seen in the middle panel of Fig.~\ref{fig:relax_ts}.  The changes in the time- and frequency-domain data after turning the PT off support that the second population of events visible in Fig.~\ref{fig:filter_vs_decay}(A) was caused by the PT.

Interestingly, once the PT was turned back on after the PT-off period, the ASD did not return to its original shape.  Although the PT frequencies became visible again, the ASD was much smaller in the low-frequency region than it was before the PT-off period, and the number of error event candidates remained low.  We note that the large PT-induced errors did not return for the duration of the measurement (several days), which extended beyond the time period shown in Fig.~\ref{fig:psds}.  The cause of this heterogeneity in device response to the PT is unknown, but it aligns with previous results suggesting that the device response to the PT varies with DR cycling or other factors \cite{kono_mechanically_2024}.  Additionally, these PT-induced correlated qubit errors were not seen in the dataset in Ref.~\onlinecite{harringtonSynchronousDetectionCosmic2024}, which was taken in DR1 with the same device design.  While the device design was the same, the chip was different, possibly resulting in small changes in the wire bonding.  This is another piece of evidence that small changes in either the device mounting or DR state can have large impacts on the presence of these PT-induced correlated errors.  These data suggest a complex relationship between the DR environment and the impact on the qubits, but it is clear that the PT is responsible for the high number of correlated qubit errors seen Section~\ref{sec:distinguishing}.

We took $T_1$ and $T_2$ measurements before the measurement period shown in Fig.~\ref{fig:psds} and after it, when the DR had returned to a steady state the following day.  Before the PT-off periods, we measured an average $T_1$ value for the qubits of 38 \unit{\micro\second} and an average $T_2$ value of 36 \unit{\micro\second}.  The following day, we measured an average $T_1$ of 44 \unit{\micro\second} and an average $T_2$ of 47 \unit{\micro\second}, demonstrating that the qubit state had not substantially worsened after the PT-off measurements.  The improvement in both $T_1$ and $T_2$ was likely due to the lack of PT-induced correlated errors, as quantified below using the PT-off data.  The $T_1$ and $T_2$ values for the 10 qubits before and after the PT-off period are given in Appendix~\ref{sec:T1_T2}.

Although $T_1$ was not measured during the PT-off period, an estimate of the effect of turning the PT off can be made using the relaxation measurements.  Reproducing the equations from Ref.~\onlinecite{harringtonSynchronousDetectionCosmic2024},
the probability of a decay in a given time period is 
\begin{equation}
    p = \frac{n_{\mathrm{decay}}}{n_{\mathrm{prep}}},
\end{equation}
where $n_{\mathrm{decay}}$ is the number of decays, and $n_{\mathrm{prep}}$ is the number of preparations into the excited state during that period.  We did not measure the excitation rate during this measurement, but we assume that the decay rate $\Gamma_{\downarrow}$ is much greater than the excitation rate $\Gamma_{\uparrow}$, which was the case for radiation-induced errors in Ref.~\onlinecite{harringtonSynchronousDetectionCosmic2024}.  Then the probability of decay for each qubit is related to $\Gamma_{\downarrow}$ by

\begin{equation}
    p = 1 - a e^{-\Gamma_{\downarrow}\Delta t},
\end{equation}
where $\Delta t = 3$ \unit{\micro\second} is the delay time between the qubit state preparation and middle of the measurement, and $a$ is a constant that accounts for preparation and measurement fidelity. 

We can relate $T_1$ to $\Gamma_{\downarrow}$ with the equation

\begin{equation}
    T_1 = \frac{1}{\Gamma_{\uparrow} + \Gamma_{\downarrow}}.
\end{equation}

\noindent To obtain a broad estimate of the effects of the PT errors on $T_1$, we assume $T_1 =\ 1/\Gamma_{\downarrow}$ and use the measured $p$ and $T_1$ values for the qubits at the beginning of the PT-on period (the red region in Fig.~\ref{fig:psds}) to calculate $a$ for each qubit.  We then assume that $a$ stays constant across the measurement period and calculate $T_1$ using $p$ in the PT-off period.  This calculation estimates that $T_1$ increased by 9 \unit{\micro s} on average when the PT was turned off, suggesting that the PT-induced errors were having a significant effect on $T_1$.

\section{\label{sec:fridge2} Measurements in a second dilution refrigerator}
\subsection{Qubit measurements}
We additionally took data with the device $\mathrm{d_1}$ in another DR, DR2, which was also precooled with a PT that had a frequency of $\approx$ 1.4 Hz.  The distribution of filter score vs. exponential decay decay lifetime is shown in Fig.~\ref{fig:filter_vs_decay}(B).  Given that PT-induced events were about 100 times more common than radiation-induced events in the DR1 dataset, it is clear that the PT-induced events were not occurring in this dataset.  We note that DR1 and DR2 were in different buildings, possibly resulting in a different ionizing radiation rate between the two measurements.  Data from DR2 is also shown in the bottom panels of Figs.~\ref{fig:relax_ts} and \ref{fig:psds}, in green.  Even when the PT was on in DR2, there were no large correlated qubit errors at the PT frequency.  

Although there were many other differences between the two DRs, we have also consistently observed higher $T_1$ and $T_2$ times for the same device in DR2.  In DR1, we measured an average $T_1$ of 38 \unit{\micro s} and an average $T_2$ of 36 \unit{\micro s}, and in DR2 we measured an average $T_1$ of 76 \unit{\micro s} and an average $T_2$ of 79 \unit{\micro s}.  Based on the improvement in $T_1$ after turning off the PT in DR1, as estimated in Section~\ref{sec:PToff}, this improvement in DR2 is likely partially, although not fully, due to the lack of PT-induced correlated errors.  Full sets of $T_1$ and $T_2$ values for the 10 qubits in DR1 and DR2 are shown  in Table \ref{tab:T1_T2}.  

We hypothesize that the lack of PT-induced correlated errors in DR2 was due to a different vibration environment in DR2 compared to DR1, which we investigated with an accelerometer, as described in the following section.

\subsection{\label{sec:accel} Accelerometer measurements}

\begin{figure}
    \centering
    \includegraphics[width=0.5\linewidth]{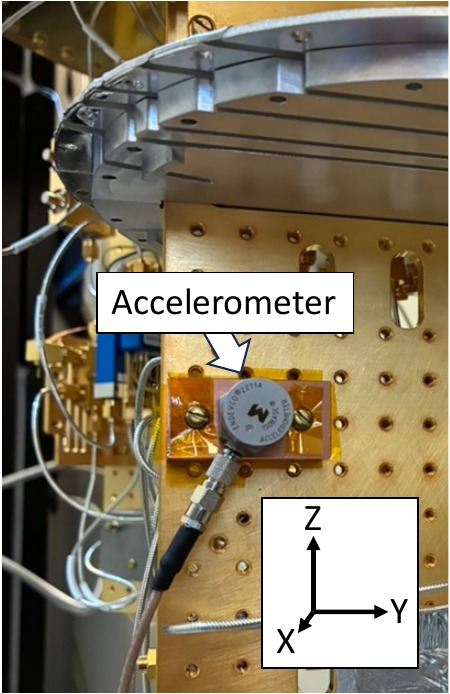}
    \caption{The accelerometer is mounted to the cold finger on the mixing chamber using a custom copper block, with the X, Y, and Z axes shown.  In this image, the accelerometer is mounted in DR2 for an X direction measurement.}
    \label{fig:accelMounting}
\end{figure}

In order to investigate the connection between the vibration environment of the two DRs and the observed correlated qubit errors, we measured the vibrations directly with an accelerometer.  

We used a \mbox{PCB 2271A} accelerometer, connected to a \mbox{PCB 422E11} charge converter and a \mbox{PCB 480E09} signal conditioner.  We took data with the accelerometer affixed to the cold finger in the X, Y, and Z orientations, as defined in Fig~\ref{fig:accelMounting}.  This figure shows the accelerometer mounted in DR2 in the X direction.  The X orientation is perpendicular to the qubit chip, whereas the Y and Z orientations are in the plane of the qubit chip.  All the data were taken at room temperature.  This allowed measurements to be taken in short succession in all three dimensions in both DRs.  The outer vacuum shield was secured, and the air was pumped out to below 10 mbar, to prevent measurement of acoustic vibrations and potential water condensation on the 4K stage.   Data were taken with the PT on for periods of 5 minutes at a 50 \unit{kHz} sampling rate in both DRs.  

\begin{figure}
    \centering
    
    \includegraphics[width=\linewidth]{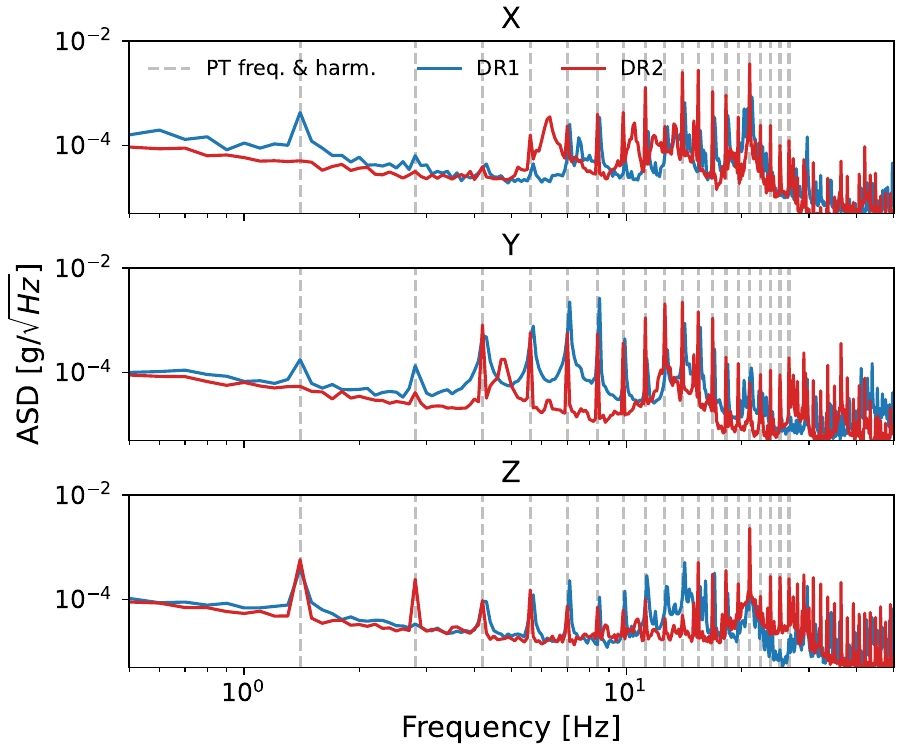}
    \caption{The amplitude spectral density in absolute acceleration units, as measured with an accelerometer, are compared for DR1 and DR2 in the X, Y, and Z directions.  DR1 shows larger peaks in X and Y at the PT frequency of $\approx$ 1.4 \unit{Hz}.}
    \label{fig:accel}
\end{figure}

Fig.~\ref{fig:accel} shows the ASD of the accelerometer data in units of $g/\sqrt{\mathrm{Hz}}$ for the X, Y, and Z directions, where $g = 9.8$ \unit{m/s^2} is a unit of acceleration.  The vertical dashed lines indicate the PT frequency and harmonics.  Although the PT frequency is visible in both DRs (at least at higher harmonics), it has a higher peak at $\approx$ 1.4 Hz in DR1 compared to DR2 in the X and Y directions.  This may explain why PT-induced errors were seen in DR1 but not in DR2, although the complexities of this coupling cannot be directly determined from this dataset.  The time series data for all of the configurations can be seen in Appendix~\ref{sec:accel_details}.

While we expect the vibration environment to change somewhat at cryogenic temperatures and with the addition of the DR thermal and magnetic shields, these measurements give a general comparison of the vibration environment in DR1 and DR2.  This difference in vibrations may be due to a variety of factors.  In DR1, the PT is rigidly affixed to the 4K plate, whereas in DR2, the PT is affixed to the 4K plate with a wide copper braid, as shown in Fig. \ref{fig:PTMounting}.  The mounting of the PT in DR2 could mitigate the vibrations.  Additionally, the DR1 plates were significantly heavier than the DR2 plates.  Other differences, such as the mounting of other devices on the DR, the vertical location of the devices, and other DR structures, could also contribute.

These results suggest a possible connection between the vibration environment of the DR and the errors shown in the previous section.  They also suggest that vibrations of the magnitude shown in Fig.~\ref{fig:accel} could cause disruptive correlated qubit errors.  Further investigations will include the operation of accelerometers at cryogenic temperatures and in coincidence with the qubits to further study the correlation.

\section{\label{sec:gapEng} Response of gap-engineered devices}
All the data described so far were taken with device $\mathrm{d_1}$.  Two additional devices, which were engineered to be less sensitive to radiation-induced correlated qubit errors, were also tested in DR1.  The filter score versus decay lifetime for device $\mathrm{d_2}$ and device $\mathrm{d_3}$ are shown in Fig.~\ref{fig:filter_vs_decay}(C) and (D), respectively.  It is immediately clear that neither radiation-induced or PT-induced events are visible.  The small number of slightly larger events seen in device $\mathrm{d_3}$ (D) compared to device $\mathrm{d_2}$ (C) were predominantly due to $T_1$ drops that only affected a single qubit at a time.

The data in Fig.~\ref{fig:filter_vs_decay}(C) and (D) were taken in the same DR cool down as the data in Fig.~\ref{fig:filter_vs_decay}(A), although the devices were not measured simultaneously.  Although the variability of the PT-induced errors, as discussed in Section~\ref{sec:distinguishing}, cannot be ruled out as the reason for the lack of PT-induced errors in devices $\mathrm{d_2}$ and $\mathrm{d_3}$, the lack of these errors in the same cool down in the same DR suggests that the gap-engineered devices are protected against these error bursts.

If the gap-engineered devices are protected against PT-induced errors, this would suggest a QP-mediated mechanism.  If PT-induced vibrations generate phonons in the device energetic enough to break Cooper pairs and increase the QP density near the JJ, then gap engineering would prevent most QPs from tunneling across it, thus mitigating PT-induced errors.  Phonons with sufficient energy for pair-breaking could be generated through the flexing of wire bonds~\cite{yelton-h65v-ttbw}, or mechanical impacts between the chip and copper housing.  If the errors were induced by electronic noise from the PT or triboelectric effects in the cables, for example, it is unlikely that they would be mitigated by gap engineering.

Although the lack of the PT-induced relaxation errors in the gap-engineered devices is very promising, recent work \cite{kurilovichCorrelatedErrorBursts2025, pinckney2026characterizationradiationinducederrorssuperconducting, nhoRecoveryDynamicsGapengineered2025} has suggested that subtler impacts of radiation can persist, even in the presence of gap engineering.  Although gap engineering prevents most relaxation errors like the ones measured in this paper, correlated phase errors can still occur from QPs present in the JJ leads \cite{kurilovichCorrelatedErrorBursts2025}.  It is possible that this effect would also occur for PT-induced errors if they are also mediated by QPs.  If we operated an accelerometer in coincidence with the gap-engineered devices, we could time align many PT impulses and average the qubit data to detect small effects, similar to the analysis in Ref.~\onlinecite{pinckney2026characterizationradiationinducederrorssuperconducting}.  This study is left for future work.

\section{\label{sec:conclusions} Conclusions}
Extensive recent research has shown that terrestrial and cosmic radiation is a source of spatiotemporally correlated qubit errors in superconducting qubits \cite{vepsalainenImpactIonizingRadiation2020, wilenCorrelatedChargeNoise2021, thorbeckTwoLevelSystemDynamicsSuperconducting2023, harringtonSynchronousDetectionCosmic2024, liDirectEvidenceCosmicrayinduced2024a, bratrudFirstMeasurementCorrelated2024a, dominicisEvaluatingRadiationImpact2024,nhoRecoveryDynamicsGapengineered2025,kurilovichCorrelatedErrorBursts2025}.  Additionally, the pulse tube cryocooler is a known source of correlated errors \cite{kono_mechanically_2024}.  In this work, we observed two types of errors in the same dataset in the same device, and we distinguished them based on their spatial, temporal, and frequency features.  Radiation-induced errors have a sharp rise time and a relatively short exponential decay lifetime.  They are also centered at varying locations on the device and do not occur at any particular frequency.  

The second type of correlated errors, which we linked to PT operation, have a longer rise time and longer decay lifetime.  They are also visible in the ASD of the data at the PT frequency.  We associated these errors with the PT by measuring the qubit devices with the PT turned off and verifying that they vanished.  We additionally measured the devices in two different DRs and found different rates of PT-induced errors.  We took measurements with an accelerometer to explore the vibration environment in both DRs and quantify the magnitude of the vibrations that likely caused the correlated errors in the first DR.  We note that these PT-induced errors can be extremely large and affect the entire device.  They also are much more common than radiation-induced errors from terrestrial and cosmogenic sources.  Therefore, mitigating them can have a significant effect on qubit coherence lifetimes.  In the second DR, we did not observe PT-induced correlated errors, likely a contributing factor to the significant improvement in $T_1$ and $T_2$ values.

Identifying the source of qubit errors allows for targeted mitigation strategies.  Radiation-induced errors can be mitigated by by operation of quantum computers in underground and shielded facilities \cite{cardani_reducing_2021, bratrudFirstMeasurementCorrelated2024a, loerAbatementIonizingRadiation2024, dominicisEvaluatingRadiationImpact2024}, as well as on-chip measures such as phonon down-conversion \cite{phonon_scattering_mitigation, iaiaPhononDownconversionSuppress2022} and gap engineering \cite{aumentadoNonequilibriumQuasiparticles2e2004, mcewenResistingHighEnergyImpact2024, nhoRecoveryDynamicsGapengineered2025,kurilovichCorrelatedErrorBursts2025}.

PT-induced errors could be mitigated by reducing vibrations.  Although the PT is necessary for the dry DRs commonly used in small scale qubit experiments, the DR configuration can have a significant effect on the rate of correlated qubit errors.  Cryogenic systems that do not use PTs could also mitigate them, but they have become less common and require additional infrastructure to run efficiently \cite{croot2025enablingtechnologiesscalablesuperconducting, Claudet2000CryoPlant}.  Vibrations are a common problem for other cryogenic devices, such as those used for fundamental physics experiments \cite{cuore_vibrations, cuore_active_noise_cancel, pirro_cryo_vibratioN_Reduction}.  Vibration isolation techniques used by those experiments could inform superconducting qubit experiment design in the future.  Additionally, gap engineering also appears effective in mitigating the effects of the PT-induced errors, as described in Section~\ref{sec:gapEng}.  Because gap engineering suppresses QP tunneling across the junction, the absence of PT-induced events in the gap-engineered devices suggests the role of QPs in mediating them.

Future investigations include studying the impacts of device mounting, vibration isolation stages, and other vibration mitigation techniques on superconducting qubits.  Additionally, cryogenic accelerometers such as the one used in this work can be operated at mixing chamber temperatures \cite{daddabbo2026commercialaccelerometersvibrationsensing}, enabling vibration measurements to be taken at the same time and location as qubit measurements.  These accelerometers could provide an independent source of information on correlated qubit errors in quantum computers.  Operating an accelerometer in coincidence with the qubits would also allow us to study whether gap-engineered devices are still subtly affected by PT-induced errors.

\begin{acknowledgments}
This research is sponsored by the U.S. Army Research Office under Award No. W911NF-23-1-0045 (Extensible and Modular Advanced Qubits), and under Air Force Contract No. FA8702-15-D-0001. The views and conclusions contained herein are those of the authors and should not be interpreted as necessarily representing the official policies or endorsements, either expressed or implied, of the U.S. Government.
\end{acknowledgments}

\appendix

\section{Details on matched filtering procedure}\label{sec:matched_filtering}
The functional form of the exponential filter used to identify error burst candidates is as follows.  We define a piecewise function $\phi_{\mathrm{exp},0}(t)$ using $\tau_0 = 5$ \unit{ms}:

 \[ 
 \phi_{\mathrm{exp},0}(t) = \begin{cases} 
      0 & t\leq 0 \\
      e^{-t/\tau_0} & t > 0
   \end{cases}
\]

To generate the array for the matched filter as shown in Figure \ref{fig:filter_vs_decay} (A, inset), the function is sampled from -0.05 to 0.05 \unit{\second} with a time between samples of 6.95 \unit{\micro\second}, for a total template length of 14388 samples.  In order to define the final template $\phi_{\mathrm{exp}}$, we subtract off the mean of the sampled array so that it sums to zero.  This makes the filter sensitive to changes in the relaxation rate.

Then, number of relaxations data is convolved with the time-reversed version of this filter using \texttt{scipy.signal.convolve}.

\section{$T_1$ and $T_2$ measurements}\label{sec:T1_T2}
Table \ref{tab:T1_T2} gives $T_1$ and $T_2$ measurements for the 10 qubits in device $\mathrm{d_1}$ in DR1 and DR2.  The first set of DR1 measurements were taken before the dataset shown in Fig.~\ref{fig:psds}, and the second set of DR1 measurements were taken after that dataset, when the PT-induced error state had changed.  The DR2 measurements were taken before the dataset shown in Fig.~\ref{fig:filter_vs_decay}(B) and Fig.~\ref{fig:psds}(B).  On average, the $T_1$ and $T_2$ values are higher in DR2 than in DR1 for the same device, with the average $T_1$ being 2 times higher and the average $T_2$ being 2.2 times higher.  This improvement is likely partly, though not fully, due to the lack of PT-induced correlated errors in DR2.

Each $T_1$ and $T_2$ coherence value is the result of a single exponential fit.

\begin{table*}[ht]
\centering
\setlength{\tabcolsep}{4pt}

\begin{tabular}{l | l | l | l l l l l l l l l l | l}
\toprule
DR & Description & Value &
Q1 & Q2 & Q3 & Q4 & Q5 & Q6 & Q7 & Q8 & Q9 & Q10 & Avg.\\
\midrule
\multirow{2}{*}{DR1} & \multirow{2}{*}{Before PT-off} & $T_1$ [\unit{\micro\second}] & 12 & 31 & 62 & 27 & 50 & 43 & 60 & 13 & 37 & 49 & 38\\
                          & & $T_2$ [\unit{\micro\second}]&  14 & 14 & 66 & 18 & 41 & 51 & 63 & 9 & 52 & 34 & 36\\
\midrule
\multirow{2}{*}{DR1} & \multirow{2}{*}{After PT-off} & $T_1$ [\unit{\micro\second}] & 9 & 56 & 55 & 62 & 68 & 36 & 48 & 30 & 44 & 32 & 44 \\
                          & & $T_2$ [\unit{\micro\second}]&   11 & 62 & 56 & 45 & 48 & 14 & 61 & 30 & 70 & 76 & 47 \\

\midrule
\multirow{2}{*}{DR2} & & $T_1$ [\unit{\micro\second}]&  24 & 18 & 109 & 110 & 74 & 92 & 107 & 101 & 34 & 88 & 76\\
                          & & $T_2$ [\unit{\micro\second}]& 40 & 7 & 131 & 27 & 160 & 84 & 157 & 75 & 42 & 66 & 79\\
\bottomrule
\end{tabular}
\caption{$T_1$ and $T_2$ values for the 10 qubits in device $\mathrm{d_1}$ measured in DR1 and DR2.  Each $T_1$ and $T_2$ coherence value is the result of a single exponential fit.}
\label{tab:T1_T2}
\end{table*}

\section{Varying the definition of the metric $A$}\label{sec:app_A}
\begin{figure}
    \centering
    \includegraphics[width=0.5\linewidth]{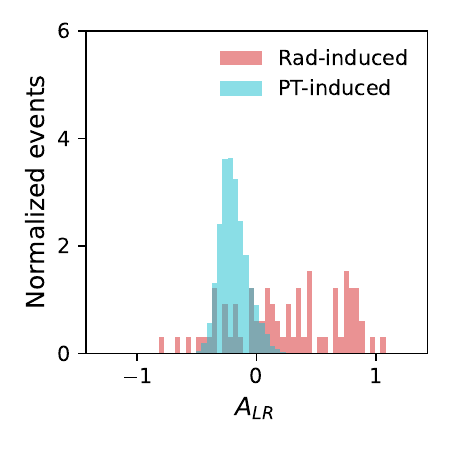}
    \caption{Histograms of the metric $A_\mathrm{LR}$ for the radiation- and PT-induced errors.  $A_\mathrm{LR}$ is the difference between the sum of the left and right qubits on the array.}
    \label{fig:A_lr}
\end{figure}

We can define the metric $A$ for any combination of qubits.  As an alternate example, we use Equation \ref{eq:A} but replace the sum over the top and bottom qubits with the sum over the left qubits (Q1-Q5) and right (Q6-Q10), respectively.  The histograms for this metric $A_\mathrm{LR}$ are shown in Figure \ref{fig:A_lr}.  We see a similar pattern as in Figure \ref{fig:localization}, with the radiation-induced events varying in $A$, while the PT-induced events have a relatively constant $A$.

\section{Pulse tube mounting}\label{sec:PTMounting}
\begin{figure}
    \centering
    \includegraphics[width=\linewidth]{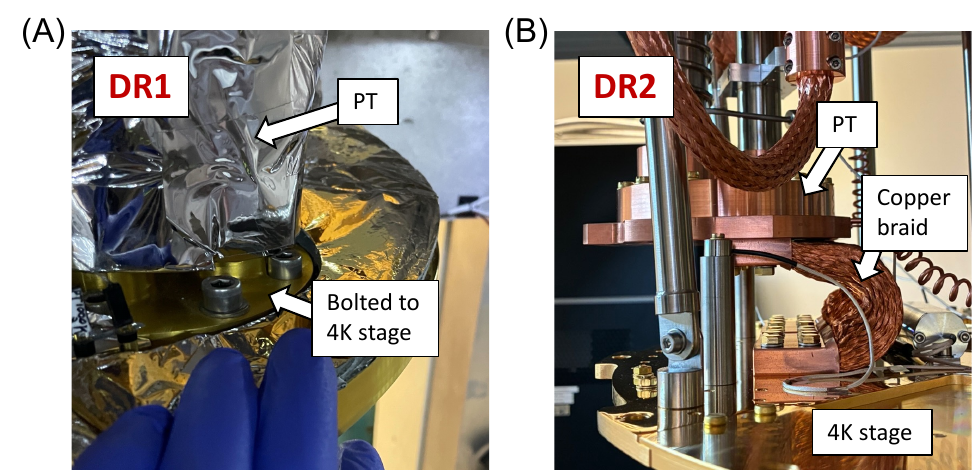}
    \caption{The mounting of the PT to the 4K stage in DR1 (A) and DR2 (B).  In DR1, the PT was rigidly bolted to the stage, whereas in DR2, the PT was attached with a wide copper braid.  In DR2, the photograph was taken from the same direction as the picture in Fig. \ref{fig:accelMounting} (i.e, X).}
    \label{fig:PTMounting}
\end{figure}

Fig. \ref{fig:PTMounting} shows the mounting of the PT to the 4K stage in DR1 (A) and DR2 (B).  In DR1, the PT was rigidly bolted to the stage, whereas in DR2, the PT was attached with a wide copper braid.  This may have contributed to the difference in vibrations in the two DRs, as discussed in Section~\ref{sec:accel}.

\section{Response of the dilution refrigerator to turning off the pulse tube}\label{sec:app_PToff}
\begin{figure}
    \centering
    \includegraphics[width=\linewidth]{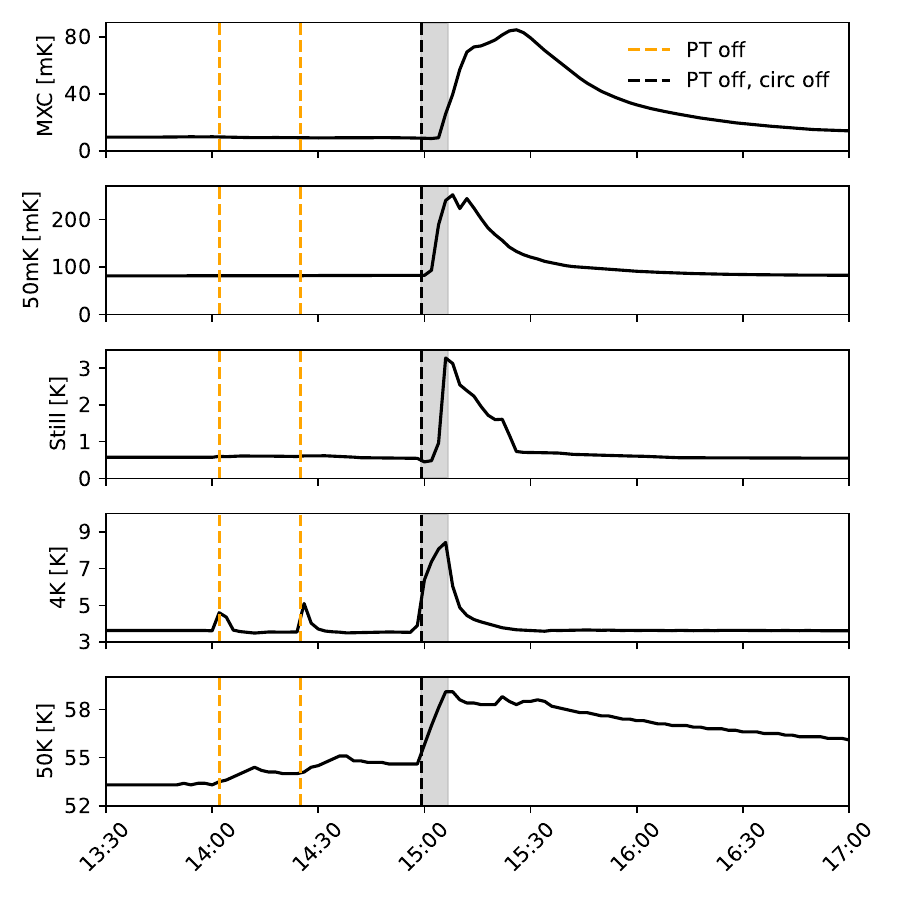}
    \caption{Response of the stages of DR1 to turning the PT off.  The orange dashed lines indicate when the PT was turned off for periods of 15-30 seconds.  The black dashed line indicates when the PT and circulation were turned off for a period of about 8 minutes, indicated by the shaded gray region.  This data corresponds to the qubit measurements in Fig.~\ref{fig:psds}.}
    \label{fig:fridge_response}
\end{figure}

This appendix gives additional details about the response of DR1 to the PT being turned off.  The temperature response of the DR stages during the measurement in Fig.~\ref{fig:psds} is shown in Fig.~\ref{fig:fridge_response}.  Initially, the PT was turned off for short periods of 15-30 seconds, marked with orange dashed lines.  This had a slight effect on the 4K and 50K stages.  Then, to take a longer PT off dataset without warming the DR, we turned off the circulation by opening a path to the He-3 dump and closing the injection line path. This prevented hot gas from entering the dilution circuit, reducing heat load on the cryostat and ideally enabling a longer measurement.  We then turned the PT off for a period of about 8 minutes, indicated by the gray shaded region, during which the temperature at all stages began to rise, with the MXC temperature peaking at around 80 mK.  We continued to take data during the entire period, as shown in Fig.~\ref{fig:psds}.  The MXC temperature took about an hour and a half to fully recover.

\begin{figure}[]
    \centering
    \includegraphics[width=\linewidth]{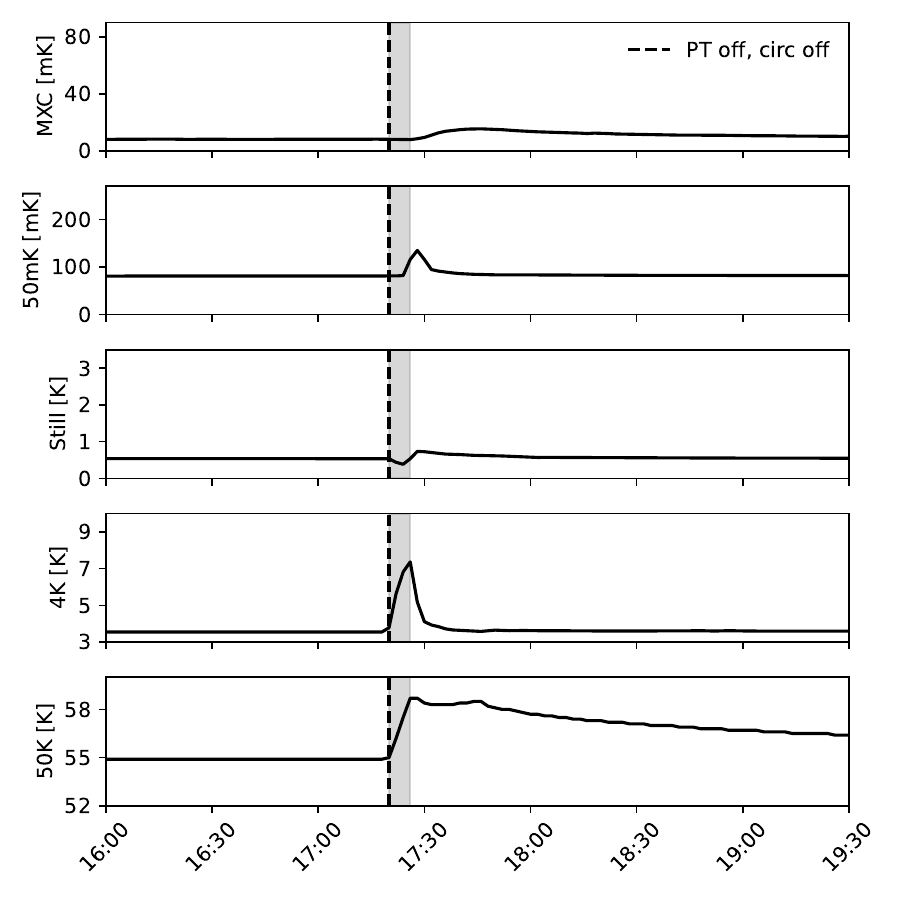}
    \caption{Response of the stages of DR1 to turning the PT off for a shorter time period.  The black dashed line indicates when the PT and circulation were turned off for a period of about 6 minutes, indicated by the shaded gray region.  This shorter PT-off period resulted in a much smaller perturbation to the DR temperatures.}
    \label{fig:fridge_response2}
\end{figure}

We also took a second set of PT-off measurements, shown in Fig.~\ref{fig:fridge_response2}, during which we turned off the PT for $\approx 6$ minutes.  By turning the PT back on before the MXC temperature began to rise, we induced significantly less perturbation in the DR temperatures.  The qubit data taken during that PT-off period are not shown in this paper, but we suggest this shorter PT-off period as the operating procedure in the future.  We also note that tests in DR2 have shown that turning off the circulation is not generally necessary for taking PT-off data.

\section{Accelerometer measurement details}\label{sec:accel_details}
Accelerometer measurements were taken with a \mbox{PCB 2271A} accelerometer.  Although this accelerometer is designed for cryogenic applications, all measurements were taken at room temperature with only the outer vacuum can (OVC) in place.  This allowed measurements to be taken in short succession in all three dimensions in both DRs.

The accelerometer was mounted to the cold finger in the mixing chamber using a custom copper block, as shown in Fig.~\ref{fig:accelMounting}.  The accelerometer was read out through an unused RF line in the DR.  It was then connected to a PCB 422E11 inline charge converter and a PCB 480E09 signal conditioner outside the DR.  The signal conditioner was set to a gain of 100.  The data were recorded using a PicoScope portable oscilloscope with a cadence of 50 kHz and a ``resolution enhancement" (averaging) of 4, corresponding to a 256 sample moving average.  The PicoScope had relatively high low-frequency noise; however, it was still sufficiently sensitive for the measurement, and it allowed for convenient use of the same data acquisition system in both DRs.

Before data were taken, the vacuum was pumped to under 10 mbar, in order to ensure that the accelerometers would not be affected by acoustic vibrations and to avoid any chance of condensation on the 4K stage.  Data were then taken with the PT on and off.  Each of the six datasets (PT on and off, for 3 axes) consisted of 30 files that were 10 seconds long, for a total of 5 minutes of data per dataset.

\begin{figure}
    \centering
    \includegraphics[width=\linewidth]{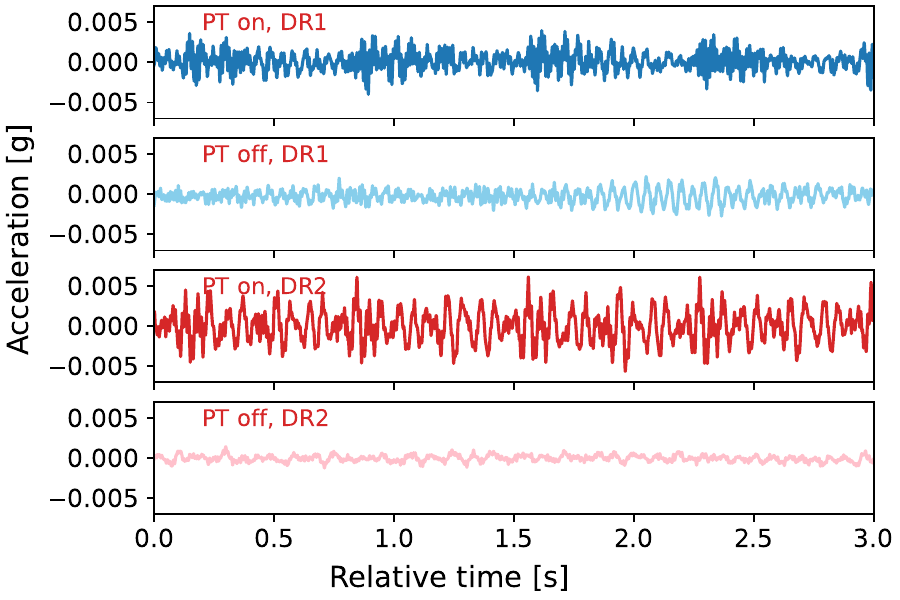}\\
    \caption{A portion of the accelerometer data in the X orientation for DR1 and DR2 with the PT on and off.}
    \label{fig:accel_ts_X}
\end{figure}

\begin{figure}
    \centering
    \includegraphics[width=\linewidth]{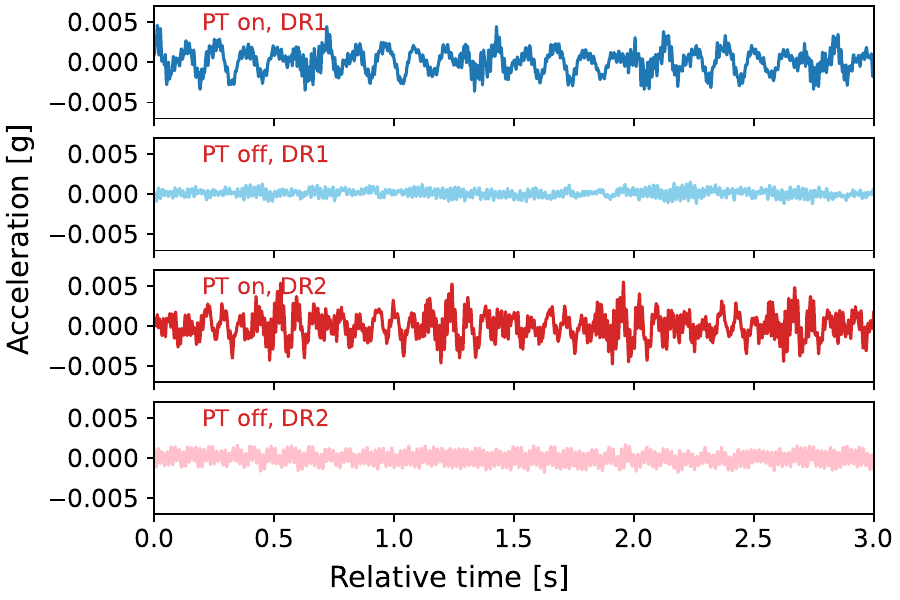}
    \caption{A portion of the accelerometer data in the Y orientation for DR1 and DR2 with the PT on and off.}
    \label{fig:accel_ts_Y}
\end{figure}

\begin{figure}
    \centering
    \includegraphics[width=\linewidth]{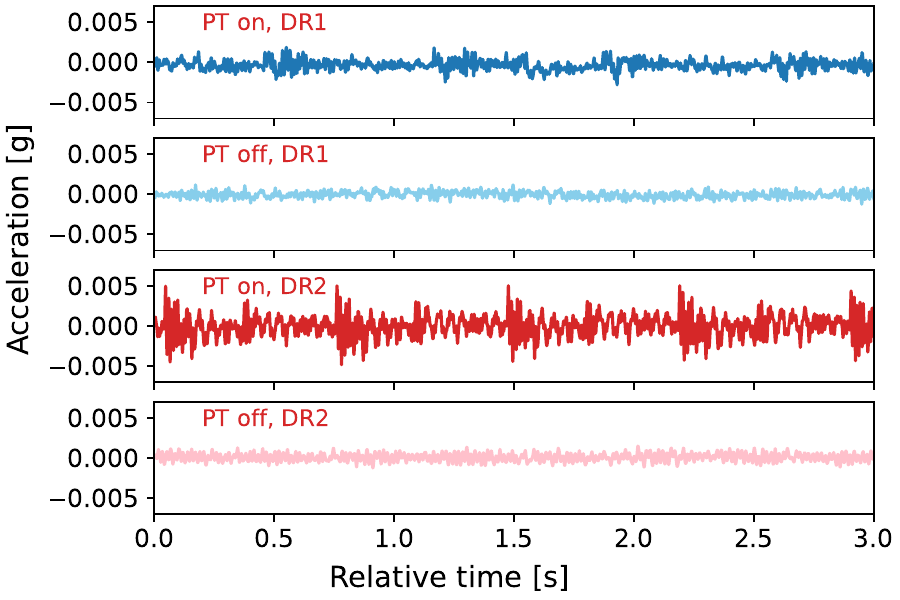}
    \caption{A portion of the accelerometer data in the Z orientation for DR1 and DR2 with the PT on and off.}
    \label{fig:accel_ts_Z}
\end{figure}

A portion of the time series data for the X, Y, and Z directions in the two DRs with the PT on and off are shown in Figs.~\ref{fig:accel_ts_X}, \ref{fig:accel_ts_Y}, and \ref{fig:accel_ts_Z}, respectively.  There are clear differences in the time series between the two DRs, providing a possible explanation for the difference in number of PT-induced correlated qubit errors.

\bibliography{correlated_error_types}

\end{document}